\documentclass[twocolumn, showpacs,preprintnumbers,amsmath,amssymb,prb,superscriptaddress]{revtex4}

\usepackage{graphicx}
\usepackage{dcolumn}
\usepackage{bm}

\begin{document}

\title {Fano Effect through Parallel-coupled Double Coulomb Islands}

\author{Haizhou Lu} \email{luhz@castu.tsinghua.edu.cn}
\affiliation{Center for Advanced Study, Tsinghua University,
Beijing 100084, China}
\author{Rong L\"{u}}
\affiliation{Center for Advanced Study, Tsinghua University,
Beijing 100084, China}
\author{Bang-fen Zhu}\email{bfzhu@castu.tsinghua.edu.cn}
\affiliation{Center for Advanced Study, Tsinghua University,
Beijing 100084, China} \affiliation{Department of Physics,
Tsinghua University, Beijing 100084, China}

\begin{abstract}
By means of the non-equilibrium Green function and equation of
motion method, the electronic transport is theoretically studied
through a parallel-coupled double quantum dots(DQD) in the
presence of the on-dot Coulomb correlation, with an emphasis put
on the quantum interference. It has been found that in the Coulomb
blockage regime, the quantum interference between the bonding and
antiboding DQD states or that between their Coulomb blockade
counterparts may result in the Fano resonance in the conductance
spectra, and the Fano peak doublet may be observed under certain
non-equilibrium condition. The possibility of manipulating the
Fano lineshape is predicted by tuning the dot-lead coupling and
magnetic flux threading the ring connecting the dots and leads.
Similar to the case without Coulomb interaction, the direction of
the asymmetric tail of Fano lineshape can be flipped by the
external field. Most importantly, by tuning the magnetic flux, the
function of four relevant states can be interchanged, giving rise
to the swap effect, which might play a key role as a qubit in the
quantum computation.

\end{abstract}


\maketitle

\section{Introduction}

Fano resonance stems from quantum interference between resonant
and nonresonant processes,\cite{Fano1961} and manifests itself in
spectra as asymmetric lineshape in a large variety of experiments.
It is known that Fano effect is a good probe for the phase
coherence for carriers in solids, in particular in a quantum dot
(QD)
system.\cite{Kouwenhoven1997,Kobayashi2002,Kobayashi2003,gores2000,zacharia2001,johnson2004}
Unlike the conventional Fano
resonance,\cite{conventional1,conventional2,conventional3,conventional4}
the Fano effect in the QD system has its advantage in that its key
parameters can be readily tuned. Suppose a discrete level inside
the QD is broadened by a factor of $\Gamma$ due to coupling to the
continua in leads, the key to realize the Fano effect in the
conductance spectra is that within $\Gamma$ centered at the
resonance level, the phase of the electron in the non-resonant
channel changes little.\cite{Yacoby1995,Schuster1997} The first
observation of the Fano lineshape in the QD system was reported by
G\"{o}res \emph{et al}.\cite{gores2000,zacharia2001} in the
single-electron-transistor experiments. Recently, K. Kobayashi
\emph{et al.} studied the magnetically and electrostatically tuned
Fano effect in a QD embedded in an Aharonov-Bohm(AB)
ring,\cite{Kobayashi2002,Kobayashi2003} and A.C. Johnson \emph{et
al}. investigated a tunable Fano interferometer consisting of a QD
coupled to a one-dimensional channel via tunneling and observed
the Coulomb-modified Fano resonances.\cite{johnson2004}

Recent experimental advances in the parallel-coupled double
quantum dots (DQD)\cite{Holleitner2001, Holleitner2002,
Holleitner2003,Blick2003,Chen2004}, in which two coupled QD's via
barrier tunneling are embedded into opposite arms of a AB ring
respectively and also coupled to two leads roughly equally ({\it
cf}  Fig.\ref{fig:model}), have inspired a number of theoretical
attempts to study the coherent transport in this
system.\cite{kang2004,guevara2003,baizhiming2004,guevara2004,Lu2005}
As a controllable two-level system, it is appealing for the
parallel-coupled DQD system to become one of promising candidates
for the quantum bit in quantum computation based on solid state
devices.\cite{Loss1998} The entangled quantum states required for
performing the quantum computation demand a high degree of phase
coherence in the system.\cite{DiVincenzo1997} Being a probe to the
phase coherence,\cite{Clerk2001} the Fano effect in the
parallel-coupled DQD system is certainly of practical importance,
if tunable, and especially if the swap effect can be manipulated.

In a real small system like quantum dots, the electron-electron
(e-e) interaction will influence the transport process. A question
naturally arises, whether the e-e interaction in the QD breaks the
Fano resonance in the DQD system or not? Till now,  most
theoretical works addressing the Fano effect in DQD ignored this
aspect by adopting a Fano-Anderson model in which the e-e
repulsion is entirely absent.
\cite{kang2004,guevara2003,baizhiming2004,guevara2004,Lu2005} The
electron correlations in the DQD have been taken into account in
coherent electronic transport through a DQD connected in a series
with electrodes,\cite{Bulka2004} however, no Fano resonance has
been found in this configuration. In this article, we intend to
investigate the Fano effect in the tunneling-coupled parallel DQD
system in the presence of finite on-dot Coulomb interactions, that
is, the Fano effect in Parallel-coupled Double Coulomb Islands.

The paper is organized as follows. After introducing the
two-impurity Anderson model with an inter-dot coupling term to
describe the parallel-coupled DQD system in Sec.\ref{sec:model},
the current through the mesoscopic system is formulated in
Sec.\ref{sec:formula} with the Green functions in the central
region and Fermi distribution in leads.\cite{meirwingreen1992} In
small $U$ cases, all spin flip processes may be ignored, within
the Hatree-Fock approximation(HFA), the current formula is reduced
to a Landauer-B\"{u}ttiker one, and the electrons with spin
$\sigma$ behave like moving in a mean field of electrons with the
opposite spin $\overline{\sigma}$. Then numerical results for the
zero temperature case are presented in Sec.\ref{sec:results}.
Owing to the on-dot repulsion and inter-dot coupling, there are
four quantum states in the DQD system relevant, centered
approximately at $\varepsilon_0-t_c$, $\varepsilon_0+t_c$,
$\varepsilon_0-t_c+U$ and $\varepsilon_0+t_c+U$, corresponding to
the bonding and antibonding DQD states and their Coulomb blockade
counterparts, respectively. It has been found that the Fano
resonance appears as a result of the quantum interference not only
between the DQD states at $\varepsilon_0-t_c$ and
$\varepsilon_0+t_c$, but also between their Coulomb blockade
counterparts. It has also been found that the direction of the
asymmetric tail of Fano lineshape can be flipped by the external
field. Most importantly, by tuning the total magnetic flux through
the AB ring, the swap effect between four resonance peaks in the
conductance spectra is predicted, which might be of potential
application as a type of C-Not gate in the quantum computation.
Finally, a brief summary is drawn and presented.

\section{Physical Model}\label{sec:model}
The total Hamiltonian is expressed as
\begin{equation}\label{hamiltonian}
\mathrm{H}=\mathrm{H}_{leads}+\mathrm{H}_{DQD}+\mathrm{H}_T.
\end{equation}

\begin{figure}[htbp]
\centering
\includegraphics[width=0.3\textwidth]{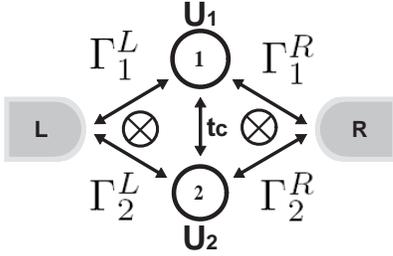}
\caption{Schematic setup of a tunneling-coupled parallel DQD
system coupled to two reservoirs.}\label{fig:model}
\end{figure}

We start with the two-impurity Anderson model for the
parallel-coupled DQD ({\it cf}  Fig.\ref{fig:model}) as
\begin{eqnarray}\label{hd1}
\mathrm{H}_{DQD}&=&\sum_{i,\sigma}\varepsilon_{i\sigma}
d^{\dagger}_{i\sigma}d_{i\sigma}
+\sum_{i}U_id^{\dagger}_{i\uparrow}d_{i\uparrow}d^{\dagger}_{i\downarrow}d_{i\downarrow}\nonumber\\
&-&t_c\sum_{\sigma}(d^{\dagger}_{1\sigma}d_{2\sigma} +h.c.),
\end{eqnarray}
where $d_{i\sigma}^{\dagger}$ $(d_{i\sigma})$ represents the
creation (annihilation) operator for the discrete state with the
energy $\varepsilon_{i\sigma}$ and spin $\sigma$ (
$\sigma=\uparrow,\downarrow$ ) in dot $i$ ( $i=1,2$), which are
coupled each other via tunneling $t_c$, and the on-dot Coulomb
repulsion is described by the second term in Eq.(\ref{hd1}).

The $\mathrm{H}_{leads}$ in Eq.(\ref{hamiltonian}) represents the
noninteracting electron gas in the left(L) and right(R) leads,
\begin{equation}
\mathrm{H}_{leads}=\sum_{k,\alpha,\sigma}\varepsilon_{k\alpha}c_{k\alpha\sigma}^{\dagger}c_{k\alpha\sigma},
\end{equation}
where $c_{k\alpha\sigma}^{\dagger}$ ($c_{k\alpha\sigma}$) is the
creation (annihilation) operator for a continuous state in the
lead $\alpha$ ($\alpha=L,R$) with energy $\varepsilon_{k\alpha}$
and spin $\sigma$.

The $\mathrm{H}_{T}$ in Eq.(\ref{hamiltonian}) represents the
tunneling coupling between the QD and lead electrons,
\begin{eqnarray}
\mathrm{H}_T=&&\sum_{k,\alpha,\sigma,i} V_{\alpha
i}d_{i\sigma}^{\dagger}c_{k{\alpha}\sigma} +  h.c.,
\end{eqnarray}
where for the sake of simplicity the tunneling matrix element
$V_{\alpha i}$ is assumed to be independent of $k$, and the phase
shift due to the total magnetic flux threading into the AB ring,
$\Phi$, is assumed to distribute evenly among 4 sections of the
DQD-AB ring, namely $V_{L1}=|V_{L1}|e^{i\frac{\phi}{4}},
V_{L2}^*=|V_{L2}|e^{i\frac{\phi}{4}},
V_{R1}^*=|V_{R1}|e^{i\frac{\phi}{4}}$, and
$V_{R2}=|V_{R2}|e^{i\frac{\phi}{4}}$. Here $\phi=
2\pi\Phi/\Phi_0$, in which the flux quantum $\Phi_0=hc/e$. In the
following calculation, we define the line-width matrix as
$\Gamma^{\alpha}_{ij}\equiv\sum_{k}V_{\alpha i}V^*_{\alpha j}2\pi
\delta(\varepsilon-\varepsilon_{k{\alpha}})$ (${\alpha}=L,R)$ and
$\mathbf{\Gamma}=\mathbf{\Gamma}^L+\mathbf{\Gamma}^R$. According
to Fig.\ref{fig:model}, the line-width matrices in the QD
representation read
\begin{equation}\label{gstrgammaL}
\mathbf{\Gamma}^L=\left(\begin{array}{cc}
 \Gamma^L_{1} &
\sqrt{\Gamma^L_{1}\Gamma^L_{2}}e^{i\frac{\phi}{2}}\\
\sqrt{\Gamma^L_{1}\Gamma^L_{2}}e^{-i\frac{\phi}{2}} &\Gamma^L_{2}
\end{array}\right),\nonumber
\end{equation}
and
\begin{equation}\label{gstrgammaR}
\mathbf{\Gamma}^R=\left(\begin{array}{cc}
 \Gamma^R_{1} &
\sqrt{\Gamma^R_{1}\Gamma^R_{2}}e^{-i\frac{\phi}{2}}\\
\sqrt{\Gamma^R_{1}\Gamma^R_{2}}e^{i\frac{\phi}{2}} &\Gamma^R_{2}
\end{array}\right),
\end{equation}
where $\Gamma^{\alpha}_i$ is short for $\Gamma^{\alpha}_{ii}$ .

\section{Current formula}\label{sec:formula}
According to Meir and Wingreen,\cite{meirwingreen1992} the general
formula for current through a mesoscopic region between
noninteracting leads is given by
\begin{eqnarray}\label{current}
J&=&\sum_{\sigma}\frac{ie}{2h}\int d\omega
\mathrm{Tr}\{(\mathbf{\Gamma}^L-\mathbf{\Gamma}^R)\mathbf{G}^{<}(\omega)\nonumber\\
&&\ \
+[f_L(\omega)\mathbf{\Gamma}^L-f_R(\omega)\mathbf{\Gamma}^R](\mathbf{G}^r(\omega)-\mathbf{G}^a(\omega))\},
\end{eqnarray}
where, $f_{L(R)}(\omega)$ is the Fermi distribution function on
the left(right) leads, $\mathbf{\Gamma}^{L(R)}$ has been given by
Eq.(\ref{gstrgammaL}), $\mathbf{G}^{r}$, $\mathbf{G}^{a}$ and
$\mathbf{G}^{<}$ are the retarded, advanced and lesser Green
functions in the DQD region, respectively.

\subsection{Retarded and Advanced Green Functions}
The retarded Green function is defined as
\begin{equation}\label{defGr}
G^r_{i\sigma,j\sigma}(t)\equiv\langle\langle
d_{i\sigma}(t)|d^{\dagger}_{j\sigma}\rangle\rangle^r\equiv-i\theta(t)\langle\{d_{i\sigma}(t),d^{\dagger}_{j\sigma}\}\rangle.
\end{equation}
Writing the equation of motion for the retarded Green function in
Fourier space $\langle\langle
d_{i\sigma}|d^{\dagger}_{j\sigma}\rangle\rangle^r_{\omega}$,\cite{Haug1996}
one arrives at
\begin{eqnarray}\label{Gijsigma}
&&(\omega-\varepsilon_{i\sigma}+\frac{i}{2}\Gamma_{ii})
\langle\langle
d_{i\sigma}|d^{\dagger}_{j\sigma}\rangle\rangle^r_{\omega}+(t_c+\frac{i}{2}\Gamma_{i\overline{i}})\langle\langle
d_{\overline{i}\sigma}|d^{\dagger}_{j\sigma}\rangle\rangle^r_{\omega}\nonumber\\&&=\delta_{ij}
+ U_i\langle\langle
d_{i\sigma}n_{i\overline{\sigma}}|d^{\dagger}_{j\sigma}\rangle\rangle^r_{\omega},
\end{eqnarray}
where, $\overline{i}=2$ if $i=1$, and vice versa. During the
derivation, we have calculated
\begin{eqnarray}\label{Galphaisigma}
\langle\langle
c_{k\alpha\sigma}|d^{\dagger}_{i\sigma}\rangle\rangle^r_{\omega}=\sum_{j=1,2}\frac{V^*_{\alpha
j}\langle\langle
d_{j\sigma}|d^{\dagger}_{i\sigma}\rangle\rangle^r_{\omega}}{\omega-\varepsilon_{k\alpha}+i0^+},
\end{eqnarray}
and defined the retarded self energy originated from the dot-lead
couplings as
\begin{equation}\label{Sigmaalphaij}
\Sigma^{r}_{i j}=\sum_{k,\alpha}\frac{V_{\alpha
i}V^*_{\alpha j}}{\omega-\varepsilon_{k\alpha}+i0^+},
\end{equation}
where $0^+$ represents an infinitesimal. In the wide-band limit,
\begin{equation}\label{Sigmaalphaij}
\Sigma^{r}_{i
j}\approx-\frac{i}{2}(\Gamma^L_{ij}+\Gamma^R_{ij})=-\frac{i}{2}\Gamma_{ij}.
\end{equation}
The equation of motion for the last term in Eq.(\ref{Gijsigma}),
$\langle\langle
d_{i\sigma}n_{i\overline{\sigma}}|d^{\dagger}_{j\sigma}\rangle\rangle^r_{\omega}$,
reads
\begin{eqnarray}\label{G1(2)}
&&(\omega-\varepsilon_{i\sigma}-U_i)\langle\langle
d_{i\sigma}n_{i\overline{\sigma}}|d^{\dagger}_{j\sigma}\rangle\rangle^r_{\omega}\nonumber\\
&=&\ \delta_{ij}\langle
n_{i\overline{\sigma}}\rangle-t_c\langle\langle
d_{\overline{i}\sigma}n_{i\overline{\sigma}}|d^{\dagger}_{j\sigma}\rangle\rangle^r_{\omega}-t_c
\langle\langle
d_{i\sigma}d^{\dagger}_{i\overline{\sigma}}d_{\overline{i}\overline{\sigma}}|d^{\dagger}_{j\sigma}\rangle\rangle^r_{\omega}\nonumber\\
&&+\ t_c\langle\langle
d_{i\sigma}d^{\dagger}_{\overline{i}\overline{\sigma}}d_{i\overline{\sigma}}|d^{\dagger}_{j\sigma}\rangle\rangle^r_{\omega}
+\sum_{k,\alpha}V_{{\alpha} i}\langle\langle c_{k\alpha
\sigma}n_{i\overline{\sigma}}|d^{\dagger}_{j\sigma}\rangle\rangle^r_{\omega}\nonumber\\
&&+\sum_{k,\alpha}V_{{\alpha} i}\langle\langle d^{\dagger}_{i\overline{\sigma}}c_{k\alpha
\overline{\sigma}}d_{i\sigma}|d^{\dagger}_{j\sigma}\rangle\rangle^r_{\omega}\nonumber\\
&&-\sum_{k,\alpha}V^*_{{\alpha} i}\langle\langle
c^{\dagger}_{k\alpha
\overline{\sigma}}d_{i\overline{\sigma}}d_{i\sigma}|d^{\dagger}_{j\sigma}\rangle\rangle^r_{\omega}.
\end{eqnarray}
To truncate the set of equations for the retarded Green function,
we adopt the HFA to the higher-order Green functions on the right
side of Eq.(\ref{G1(2)}) and have
\begin{eqnarray}\label{hartree}
\langle\langle
d_{\overline{i}\sigma}n_{i\overline{\sigma}}|d^{\dagger}_{j\sigma}\rangle\rangle^r_{\omega}
&\cong&\langle n_{i\overline{\sigma}}\rangle\langle\langle
d_{\overline{i}\sigma}|d^{\dagger}_{j\sigma}\rangle\rangle^r_{\omega},\nonumber\\
\langle\langle
d_{i\sigma}d^{\dagger}_{i\overline{\sigma}}d_{\overline{i}\overline{\sigma}}|d^{\dagger}_{j\sigma}\rangle\rangle^r_{\omega}
&\cong&\langle
d^{\dagger}_{i\overline{\sigma}}d_{\overline{i}\overline{\sigma}}
\rangle\langle\langle
d_{i\sigma}|d^{\dagger}_{j\sigma}\rangle\rangle^r_{\omega},\nonumber\\
\langle\langle c_{k\alpha
\sigma}n_{i\overline{\sigma}}|d^{\dagger}_{j\sigma}\rangle\rangle^r_{\omega}
&\cong&\langle n_{i\overline{\sigma}}\rangle\langle\langle
c_{k\alpha \sigma}|d^{\dagger}_{j\sigma}\rangle\rangle^r_{\omega},\nonumber\\
\langle\langle d^{\dagger}_{i\overline{\sigma}}c_{k\alpha
\overline{\sigma}}d_{i\sigma}|d^{\dagger}_{j\sigma}\rangle\rangle^r_{\omega}&\cong&
 \langle d^{\dagger}_{i\overline{\sigma}}c_{k\alpha
\overline{\sigma}}\rangle \langle\langle
d_{i\sigma}|d^{\dagger}_{j\sigma}\rangle\rangle^r_{\omega}.
\end{eqnarray}
Thus obtained retarded Green function is expressed in a compact
form as
\begin{equation}\label{G}
\mathbf{G}^r(\omega)=[1-\mathbf{g}^r(\omega)\mathbf{\Sigma}^r]^{-1}\mathbf{g}^r(\omega),
\end{equation}
in which $\mathbf{\Sigma}^r$ is given by Eqs. (\ref{gstrgammaR})
and (\ref{Sigmaalphaij}), and $\mathbf{g}^r(\omega)$ is the Green
function for the isolated DQD. It is convenient to express the
inverse of $\mathbf{g}^r(\omega)$ as
\begin{eqnarray}
[\mathbf{g}^{r}(\omega)^{-1}]_{ii}&=&\frac{(\omega-\varepsilon_{i\sigma})(\omega-\varepsilon_{i\sigma}-U_i)}{\omega-\varepsilon_{i\sigma}-U_i+U_i\langle
n_{i\overline{\sigma}}\rangle}\nonumber\\
&&+\frac{U_i t_c[\langle
d^{\dagger}_{i\overline{\sigma}}d_{\overline{i}\overline{\sigma}}\rangle
- \langle
d^{\dagger}_{\overline{i}\overline{\sigma}}d_{i\overline{\sigma}}\rangle]}{\omega-\varepsilon_{i\sigma}-U_i+U_i\langle
n_{i\overline{\sigma}}\rangle},\nonumber
\end{eqnarray}
and
\begin{equation}\label{G-1ii}
[\mathbf{g}^r(\omega)^{-1}]_{i\overline{i}}=t_c.
\end{equation}
Here, the expectation values of $\langle
n_{\i\overline{\sigma}}\rangle$ and $\langle
d^{\dagger}_{i\overline{\sigma}}d_{\overline{i}\overline{\sigma}}\rangle$
can be calculated self-consistently by taking advantage of the
definition of the lesser Green function
\begin{equation}\label{nisigmabar}
\langle n_{i\overline{\sigma}} \rangle = -i\int^{\infty}_{-\infty}
\frac{d\omega}{2\pi}
G^{<}_{i\overline{\sigma},i\overline{\sigma}}(\omega),
\end{equation}
and
\begin{equation}\label{ddaggerdsigmabar}
\langle
d^{\dagger}_{i\overline{\sigma}}d_{\overline{i}\overline{\sigma}}\rangle
= -i\int^{\infty}_{-\infty}\frac{d\omega}{2\pi}
G^{<}_{\overline{i}\overline{\sigma},i\overline{\sigma}}(\omega).
\end{equation}

The derivation of advanced Green function $\mathbf{G}^a$ follows
the same procedure, and its expression is the Hermite conjugate of
the retarded Green function.

\subsection{Lesser Green Function}
The lesser Green function is defined by
\begin{equation}\label{defG<}
G^{<}_{i\sigma,j\sigma}(t)\equiv\langle\langle
d_{i\sigma}(t)|d^{\dagger}_{j\sigma}\rangle\rangle ^<\equiv
i\langle d^{\dagger}_{j\sigma}d_{i\sigma}(t)\rangle.
\end{equation}
The equation of motion for the lesser Green function in the
Fourier space follows as
\begin{eqnarray}\label{eom11}
&&(\omega-\varepsilon_{i\sigma}) \langle\langle
d_{i\sigma}|d^{\dagger}_{j\sigma}\rangle\rangle ^<_{\omega}+t_c
\langle\langle
d_{\overline{i}\sigma}|d^{\dagger}_{j\sigma}\rangle\rangle ^<_{\omega}\nonumber\\
&=&\sum_{k,\alpha}V_{\alpha i} \langle\langle
c_{k\alpha\sigma}|d^{\dagger}_{j\sigma}\rangle\rangle
^<_{\omega}  +U_i\langle\langle
d_{i\sigma}n_{i\overline{\sigma}}|d^{\dagger}_{j\sigma}\rangle\rangle^<_{\omega}.
\end{eqnarray}
The first term on the right hand side of Eq.(\ref{eom11}) can be
achieved with the analytic continuation\cite{Langreth1976} as
follows
\begin{eqnarray}\label{eom7}
&&\langle\langle
c_{k\alpha\sigma}|d^{\dagger}_{j\sigma}\rangle\rangle
^<_{\omega}=
\sum_{l=1,2}V^*_{\alpha l}\nonumber\\
&&\times[g^r_{k\alpha}(\omega) \langle\langle
d_{l\sigma}|d^{\dagger}_{j\sigma}\rangle\rangle ^<_{\omega}
+g^<_{k\alpha}(\omega) \langle\langle
d_{l\sigma}|d^{\dagger}_{j\sigma}\rangle\rangle ^a_{\omega}],
\end{eqnarray}
where $g^{r(a)}_{k\alpha}$ is the Green function for the
noninteracting electrons in the leads,
\begin{eqnarray}\label{grglesser}
g^r_{k\alpha}(\omega)&=&\frac{1}{\omega-\varepsilon_{k\alpha}+i0^+},\nonumber\\
g^<_{k\alpha}(\omega)&=&if_{\alpha}(\omega)2\pi\delta(\omega-\varepsilon_{k\alpha}).
\end{eqnarray}
The equation of motion for the last term on the right hand side of
Eq.(\ref{eom11}) has almost the same structure with
Eq.(\ref{G1(2)}) except for term $\delta_{ij}\langle
n_{i\overline{\sigma}}\rangle$. Under the Hartree-Fock
approximation, by using the expressions similar to
Eq.(\ref{hartree}), we eventually arrive at
\begin{eqnarray}\label{G1(2)HF2}
&&(\omega-\varepsilon_{i\sigma}-U_i)\langle\langle
d_{i\sigma}n_{i\overline{\sigma}}|d^{\dagger}_{j\sigma}\rangle\rangle^<_{\omega}\nonumber\\
&=&-\ t_c\langle n_{i\overline{\sigma}}\rangle \langle\langle
d_{\overline{i}\sigma}|d^{\dagger}_{j\sigma}\rangle\rangle^<_{\omega}
-t_c\langle
d^{\dagger}_{i\overline{\sigma}}d_{\overline{i}\overline{\sigma}}\rangle
\langle\langle d_{i\sigma}
|d^{\dagger}_{j\sigma}\rangle\rangle^<_{\omega}\nonumber\\
&&+\ t_c\langle
d^{\dagger}_{\overline{i}\overline{\sigma}}d_{i\overline{\sigma}}\rangle\langle\langle
d_{i\sigma}
|d^{\dagger}_{j\sigma}\rangle\rangle^<_{\omega}\nonumber\\
&&+\sum_{l=1,2}\langle
n_{i\overline{\sigma}}\rangle[-\frac{i}{2}\Gamma_{il}
\langle\langle
d_{l\sigma}|d^{\dagger}_{j\sigma}\rangle\rangle^<_{\omega}\nonumber\\
&&\ \ \ \ \ \ \ \ \ \ \ \ \ \ \
+i(f_L\Gamma^L_{il}+f_R\Gamma^R_{il}) \langle\langle
d_{l\sigma}|d^{\dagger}_{j\sigma}\rangle\rangle^a_{\omega}
].
\end{eqnarray}
Inserting Eqs.~(\ref{eom7}) and (\ref{G1(2)HF2}) into
Eq.~(\ref{eom11}), the expression of $\mathbf{G}^<$ can be simply
cast into
\begin{equation}\label{Gless}
\mathbf{G}^<=\mathbf{G}^r\mathbf{\Sigma}^<\mathbf{G}^a,
\end{equation}
where
$\mathbf{\Sigma}^<=i(f_L\mathbf{\Gamma}^L+f_R\mathbf{\Gamma}^R)$,
indicating that, with the second-order HFA, the self energies due
to the dot-lead coupling are separated from those resulting from
Coulomb interaction.

\subsection{Current Formula}
Generally,
$\mathbf{G}^r-\mathbf{G}^a=\mathbf{G}^r(\mathbf{\Sigma}^r-\mathbf{\Sigma}^a)\mathbf{G}^a$.
With the discussion above, the Eq.(\ref{current}) of current is
simplified to\cite{meirwingreen1992}
\begin{equation}\label{currentnoninteracting}
J=\sum_{\sigma}\frac{e}{h}\int d\omega
[f_L(\omega)-f_R(\omega)]\mathrm{Tr}[\mathbf{G}^a(\omega)
\mathbf{\Gamma}^R \mathbf{G}^r(\omega) \mathbf{\Gamma}^L ],
\end{equation}
where
$f_{L(R)}=1/\{\mathrm{exp}[{(\omega-\mu\mp\frac{eV}{2})/k_BT}]+1\}$,
and $V$ is the applied bias voltage. This current expression
reduces to a usual Landauer-B\"{u}ttiker formula for the
noninteracting case, implying that, with up to the second order of
the HFA, the system can be effectively described by a
single-particle picture. In this context, the effect of electrons
with spin $\overline{\sigma}$ on the motion of the electron with
spin $\sigma$ behaves like a background, and the coherent
tunneling process takes place between electrons with the same
spin.

\section{Numerical results at zero temperature}\label{sec:results}

The parameters for our model calculation are taken as follows. For
simplicity, we assume
$\varepsilon_{1\sigma}=\varepsilon_{2\sigma}=\varepsilon_0=0$.
Throughout this work, we consider the case where the on-dot
charging energy $U$ is much larger than the interdot coupling
$t_c$ which is of the order of unity, thus we take $U_1=U_2=U=4$.
As pointed out in Ref~.\onlinecite{guevara2003}, in the
parallel-coupled geometry, if $\Gamma^{L}_1=\Gamma^{L}_2$ and
$\Gamma^R_1=\Gamma^R_2$, the antibonding state could be decoupled
entirely from the leads. Our calculation verifies that this
situation is retained when the on-dot Coulomb repulsion is taken
into account. To avoid it, in the following we only choose two
configurations: (1) $\Gamma^L_1=\Gamma^R_2>\Gamma^L_2=\Gamma^R_1$,
and (2) $\Gamma^L_1=\Gamma^R_1>\Gamma^L_2=\Gamma^R_2$.

It should be pointed out that the numerical results presented in
this Section are valid only at temperature above the Kondo
temperature, though are calculated at zero temperature, because all
the spin flip processes have been neglected as mentioned above.

\subsection{Equilibrium process}

We are particularly interested in how the states in the DQD region
are modified by the on-dot Coulomb repulsion, and how this
modification influences transport properties. First, let us
estimate the eigenstate and eigenenergy of the isolate DQD system.

In the case of only one electron in the DQD system, due to the
inter-dot coupling, the DQD states are the linear combination of
the states in two dots, thus the formed bonding and antibonding
states are associated with energy at $\varepsilon_0-t_c$ and
$\varepsilon_0+t_c$, respectively. Hence, the one-electron ground
state of the DQD is the bonding state with eigenenergy
$\varepsilon_0-t_c$.

When the DQD contains two electrons, 6 possible states in the
system include: $|\uparrow\rangle_1|\uparrow\rangle_2$,
$|\downarrow\rangle_1|\downarrow\rangle_2$,
$|\uparrow\rangle_1|\downarrow\rangle_2$,
$|\downarrow\rangle_1|\uparrow\rangle_2$,
$|\uparrow\downarrow\rangle_1|0\rangle_2$,
$|0\rangle_1|\uparrow\downarrow\rangle_2$. The ground and excited
two-electron-states are determined by directly diagonalizing the
matrix representation of $\mathrm{H}_{DQD}$, spanned in the
Hilbert space by these 6 states. Thus the two-electron ground
state is associated with an energy at
$2\varepsilon_0+\frac{1}{2}(U-\sqrt{U^2+16t_c^2})\approx2\varepsilon_0-4t_c^2/U\sim2\varepsilon_0$.
Since the intra-dot Coulomb interaction produces effective
charging energy on the bonding and antibonding states, it is
expected for the two-electron ground state that the electrons tend
to distribute themselves evenly throughout the DQD structure to
avoid the charging energy.

Because an extra charging energy has to be consumed to add the
third and fourth electron into the system, then the ground states
energy is approximately $3\varepsilon_0+U-t_c$ and
$4\varepsilon_0+2U$, respectively.

\begin{figure}[htbp]
\centering
\includegraphics[width=0.5\textwidth]{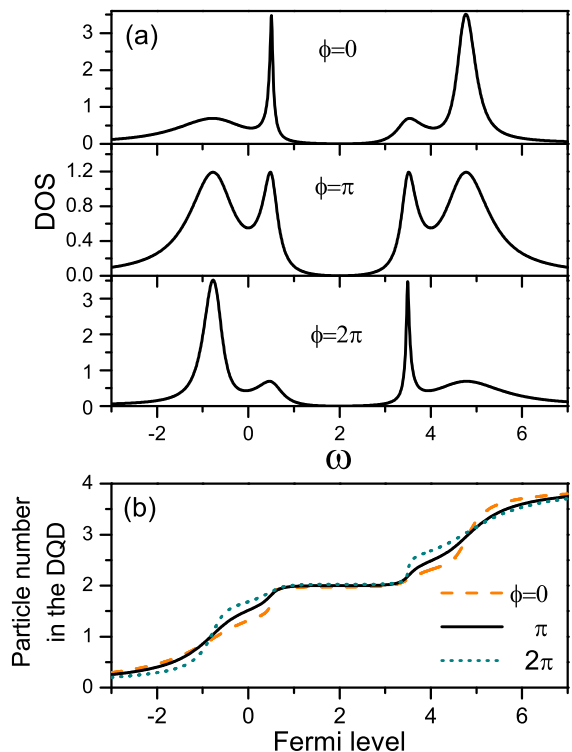}
\caption{(Color online) (a)The local density of states and (b) the
total particle number in the DQD structure for different magnetic
phase parameter $\phi$. A magnetic flux threading into the AB-ring
swaps the effective couplings of the DQD states as well as their
Coulomb counterparts to the leads. The parameters taken for
calculation are: $\varepsilon_{1\sigma}=\varepsilon_{2\sigma}=0$,
$t_c=1$, $U_1=U_2=4$, $\Gamma^L_1=\Gamma^R_2=1$ and $
\Gamma^L_2=\Gamma^R_1=0.15$. Same results are obtained when the
dot-lead coupling is changed to $\Gamma^L_1=\Gamma^R_1=1$ and
$\Gamma^L_2=\Gamma^R_2=0.15$.}\label{fig:dos}
\end{figure}

When an isolate DQD is connected to two leads, the couplings
between the DQD levels and two leads result in broadening of the
discrete levels and forming bands. The local density of states
(DOS), defined as the imaginary part of the retarded Green
function
$\rho_{\sigma}=-\frac{1}{\pi}\sum_{i=1,2}\mathrm{Im}G^r_{i\sigma,i\sigma},$
have been calculated for three different magnetic flux $\phi$.
Fig.~\ref{fig:dos}(a) reveals that the lineshape of the local DOS
critically depends on magnetic flux $\phi$ through the modulation
of the effective coupling between the DQD states and leads, which
can be tuned by both the dot-lead coupling strength and the total
magnetic flux. Fig.~\ref{fig:dos}(a) also shows the band width
variation versus the magnetic flux: The broadening of the bonding
DQD state is always accompanied by the shrinking of the
antibonding DQD state; same for their Coulomb counterparts, though
the lineshape of the DQD states and their Coulomb counterparts are
somewhat different. It is also noted from Fig.~\ref{fig:dos}(a)
that, however, unlike without Coulomb repulsion case\cite{Lu2005}
where the total width of the bonding and antibonding bands is an
invariant because the self-energy is solely determined by the
DQD-leads coupling; with the on-dot Coulomb correlation into
consideration, an additional self-energy due to Coulomb repulsion
plays a role, then the total band width in general depends on the
magnetic flux $\phi$ to some extent.

According to Eq.(\ref{currentnoninteracting}), the differential
conductance at equilibrium is defined as
$\mathcal{G}(\mu)=\frac{\partial J}{\partial V}|_{V\rightarrow
0}$, which at zero temperature reads
\begin{equation}\label{conductancenoninteracting}
\mathcal{G}(\mu)=\sum_{\sigma}\frac{e^2}{h}\mathrm{Tr}[\mathbf{G}^a(\mu)
\mathbf{\Gamma}^R \mathbf{G}^r(\mu) \mathbf{\Gamma}^L ],
\end{equation}
where $\mu$ is the Fermi level. When Fermi level varies, the
occupation number of electrons in the DQD region is changed
correspondingly. As shown in Fig.\ref{fig:dos}(b), the integer
number of electrons confined to the DQD region occurs
approximately at the following energies: $\varepsilon_0-t_c,\
\varepsilon_0+t_c,\ \varepsilon_0 + U - t_c,\ \varepsilon_0 + U +
t_c$.\cite{Datta1994,Bulka2004}

\begin{figure}[htbp]
\centering
\includegraphics[width=0.5\textwidth]{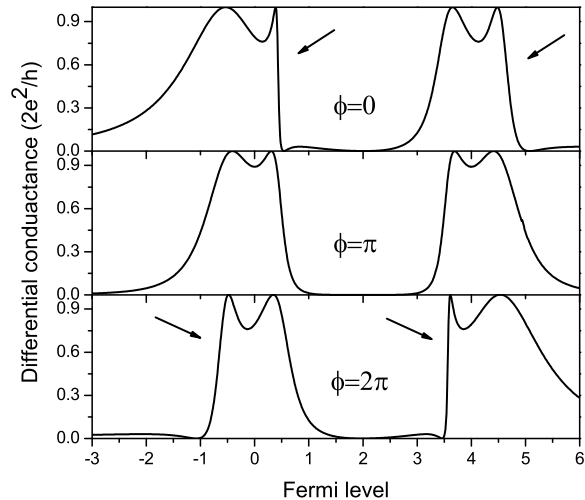}
\caption{The differential conductance for configuration 1 as a
function of Fermi level. The Fano resonances are marked with
arrows. The parameters taken are:
$\varepsilon_{1\sigma}=\varepsilon_{2\sigma}=0$, $t_c=0.8$,
$\Gamma^L_1=\Gamma^R_2=1$, $\Gamma^R_1=\Gamma^L_2=0.15$, and
$U_1=U_2=4$.}\label{fig:swapc1gp15}
\end{figure}

\begin{figure}[htbp]
\centering
\includegraphics[width=0.5\textwidth]{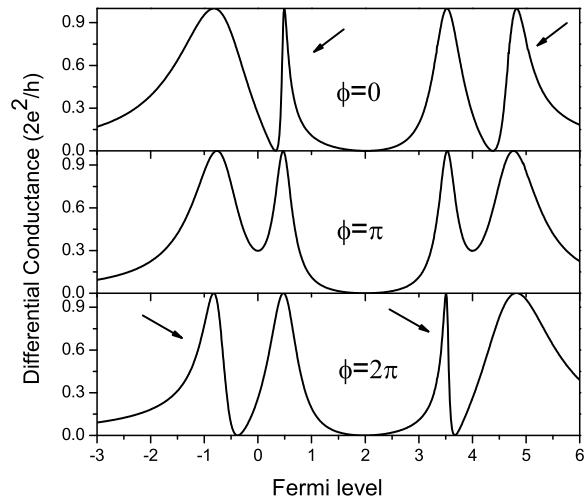}
\caption{The differential conductance for configuration 2 as a
function of Fermi level. The Fano resonances are marked with
arrows. The parameters taken are the same to the figure above
except for  $t_c=1$, $\Gamma^L_1=\Gamma^L_2=1$, and
$\Gamma^R_1=\Gamma^R_2=0.1$.}\label{fig:swapc2gp15}
\end{figure}

Figs.~\ref{fig:swapc1gp15} and \ref{fig:swapc2gp15} show the
differential conductance spectra as functions of the Fermi level
(or equivalently, of the dot level $\varepsilon_0$) in two
different configurations. The peaks marked with arrows represent
the peaks of the Fano-type, whose lineshape is asymmetric compared
with the symmetric Lorentzian at the same spectra. Four Lorentzian
peaks appear in the conductance spectra when $\phi=\pi$; while
there are only two Lorentzian and two Fano peaks for $\phi=0$. The
spectra for $\phi=2\pi$ has a mirror symmetry with that for
$\phi=0$.

The peak positions at the DOS and conductance spectra we obtained
are in good agreement with those in Refs.~\onlinecite{Datta1994}
and \onlinecite{Bulka2004}, in which two dots coupled in series
were considered; however, no Fano effect was observed in the
in-series-coupled DQD system. It is then necessary to explain why
the Fano interference occurs in the parallel-coupled DQD system,
while absents from the in-series DQD system.

As pointed out in our recent article,\cite{Lu2005} in general, the
lead states couple to the bonding and antibonding levels with
different strengths, leading to quite different broadening. If the
band width of the strongly-coupled level covers the weakly-coupled
band and the phase shift for states in the strongly-coupled
channel is negligibly small when across the weakly-coupled level,
then the Fano interference will occur, in which the weakly-coupled
channel acts as a Breit-Wigner scatter in the resonant tunneling
process while the strongly-coupled channel can be regarded as a
non-resonant one. For example, the transmission amplitude of an
electron resonantly traversing through the a weakly-coupled energy
level at $\epsilon_0$ can be described as\cite{Clerk2001}
\begin{equation}\label{propagator}
t_R\frac{\Gamma}{\omega-\epsilon_0+i\Gamma},
\end{equation}
which implies a phase shift by $\pi$ for the transmission
amplitude over an energy range of $\Gamma$ around this level, and
$t_R$ is a coefficient. On the other hand, the transmission
amplitude for states in the strongly-coupled channel can be
approximated as $t_N e^{i\phi_N}$, in which the phase $\phi_N$
varies little when across the narrow band centered at
$\epsilon_0$, the interference between these two channels yields
the Fano lineshape around the weakly-coupled level $\epsilon_0$ as
\begin{equation}\label{Fanolineshape}
|t_R\frac{\Gamma}{\omega-\varepsilon_0+i\Gamma}+t_N
e^{i\phi_N}|^2=t^2_N\frac{|\widetilde{\varepsilon}+q|^2}{\widetilde{\varepsilon}^2+1},
\end{equation}
where the detuning
$\widetilde{\varepsilon}=(\omega-\varepsilon_0)/\Gamma$, and the
asymmetric factor $q=i+ t_R e^{-i\phi_N}/t_N $. It should be noted
that in the present system, the resonant or non-resonant channel
is not fixed, on the contrary, it could be any one of the four
channels, as long as the required magnetic flux  as well as Fermi
level are satisfied.

Notice also that when the dot-lead coupling strength is adjusted
such that the configuration 1 (Fig.\ref{fig:swapc1gp15}) is
transformed into the configuration 2 (Fig.\ref{fig:swapc2gp15}),
the tail direction of the Fano peak is flipped. The reason behind
this behavior is that, compared to configuration 2, an extra flux
of $\pi$ threads into the loop in the configuration
1,\cite{Lu2005} then the Fano lineshape in configuration 1 is just
opposite to that in configuration 2. Namely, if two channels
interfere with each other constructively in configuration 1, they
will interfere destructively in configuration 2, and vice versa.
Because of different inter-dot coupling $t_c$, the splitting
between the bonding and antibonding bands in
Figs.~\ref{fig:swapc1gp15} and \ref{fig:swapc2gp15} is different,
resulting in different Fano lineshapes in these two figures.

As for the in-series DQD geometry, the bonding and antibonding
states are coupled to the leads with equal strength since one dot
has to couple to the left or right leads via the other dot. Though
the quantum interference can occur in this geometry as well, one
DQD state can not cover or be covered by the spectrum of the other
state to give rise to the Fano resonance.

\subsection{Out of Equilibrium}\label{sec:nonequilibrium}
\begin{figure}[htbp]
\centering
\includegraphics[width=0.5\textwidth]{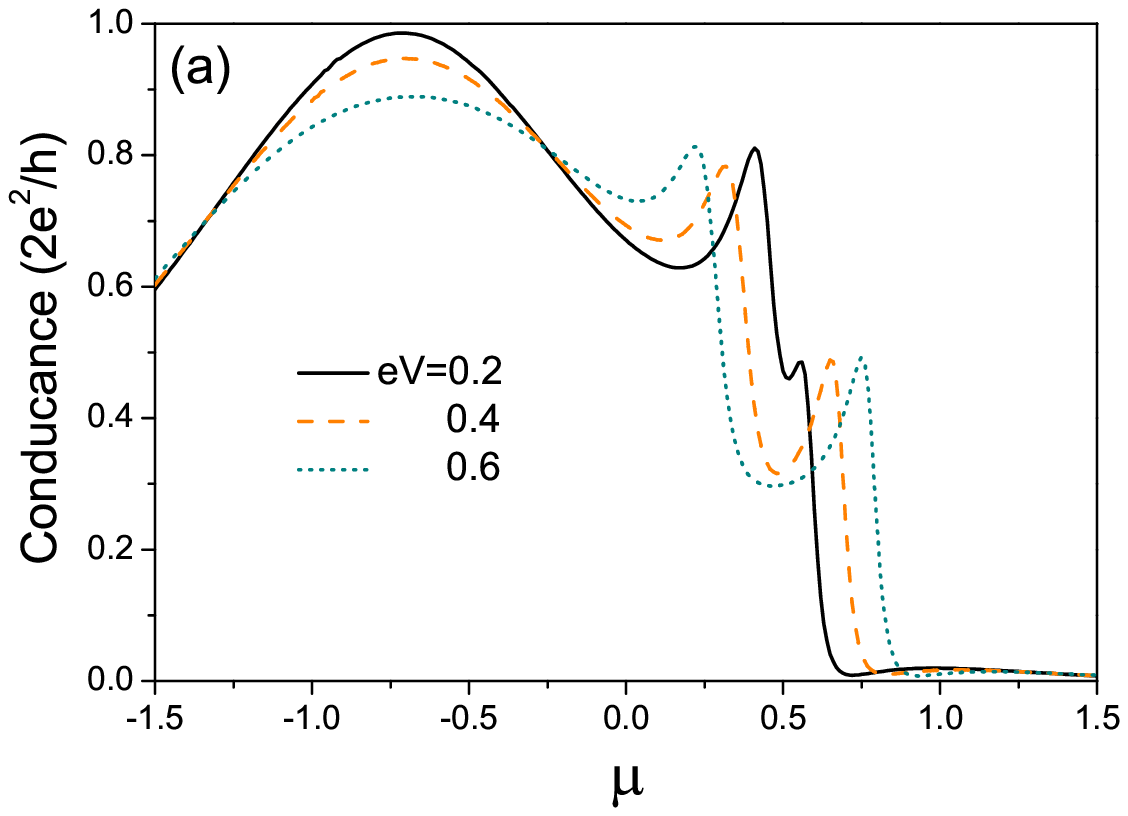}
\includegraphics[width=0.5\textwidth]{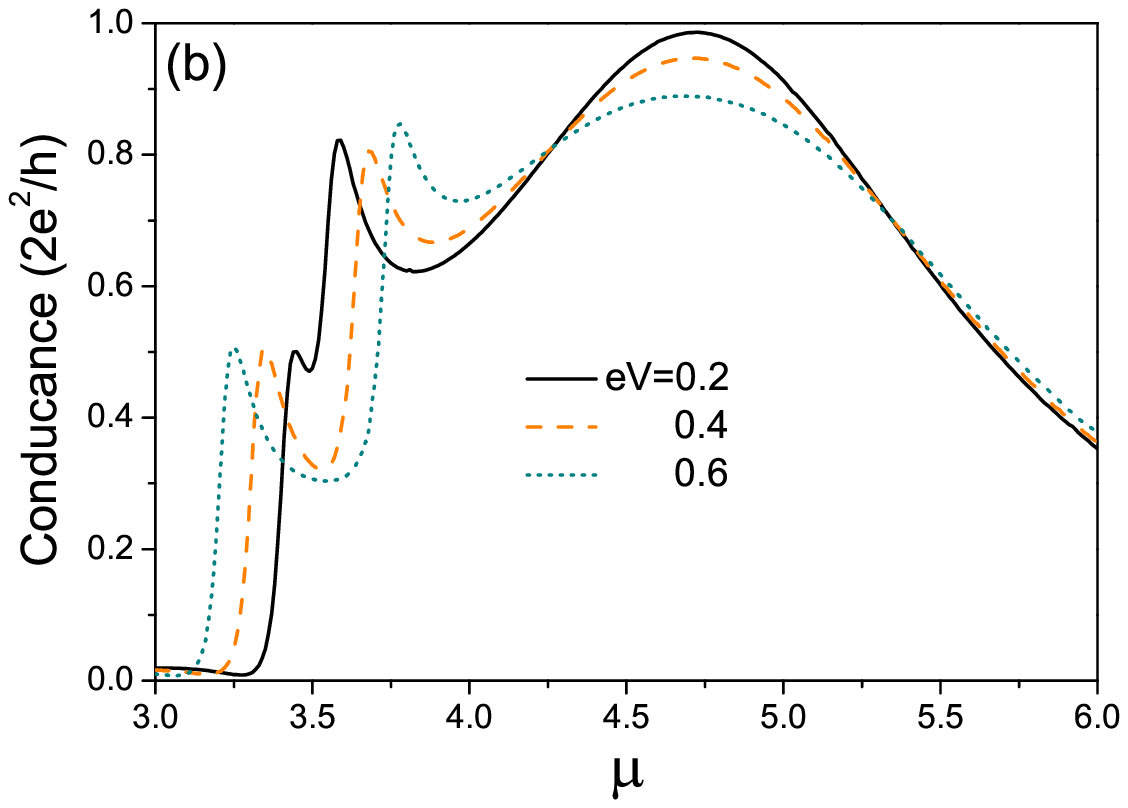}
\caption{(Color online) The conductance as functions of the
average of the left and right Fermi levels, $\mu$, for different
bias voltages in configuration 1. (a) The resonances at
$\varepsilon_0-t_c$ and $\varepsilon_0+t_c$ for $\phi=0$. (b) The
resonances at $\varepsilon_0+U-t_c$ and $\varepsilon_0+U+t_c$ for
$\phi=2\pi$.}\label{fig:split}
\end{figure}

In the presence of a finite bias voltage $V$, the differential
conductance Eq.~(\ref{currentnoninteracting}) becomes
\begin{equation}\label{nonequilibriumconductance}
     \mathcal{G}(\mu)=\sum_{\sigma}\frac{e^2}{2h}\sum_{\alpha=L,R}\mathrm{Tr}
    [\mathbf{G}^a(\mu_{\alpha})\mathbf{\Gamma}^R\mathbf{G}^r(\mu_{\alpha})\mathbf{\Gamma}^L],
\end{equation}
where $\mu_{L(R)}=\mu \pm \frac{eV}{2}$, and
$\mu=(\mu_L+\mu_R)/2$. Varying $\mu$ is equivalent to adjusting
the DQD energy levels reversely.
Eq.~(\ref{nonequilibriumconductance}) indicates that the left and
right leads contribute to the differential conductance separately.
Thus, the Fano peak doublet in the conductance spectra can be
expected, when  either of two Fermi levels is aligned with the
weakly-coupled DQD level under certain circumstance. More
specifically, when the system is driven by a bias $V$, if $eV$ is
comparable with or greater than the width of the weakly-coupled
level, but less than the width of the strongly-coupled level, the
original Fano peak in the conductance spectra may be split. In
Fig.\ref{fig:split}, two examples are presented: (a) The
conductance spectra around
 $\varepsilon_0-t_c$ and $\varepsilon_0+t_c$ in
the absence of magnetic flux, and (b) the spectra around
$\varepsilon_0+U-t_c$ and $\varepsilon_0+U+t_c$ in the presence of
a $2\pi$ flux. In both cases, three values of $eV$ locate within a
region, which is comparable with the width of the weakly-coupled
states and less than the width of the strongly-coupled states. It
is obvious from the Figure that the splitting of the Fano doublet
is proportional to the applied bias, which, together with the
Lorentzian peak, makes a step-like lineshape in the conductance
spectra.

\section{Conclusions}
In summary, within the Keldysh non-equilibrium Green function
formalism and by using the equation of motion method in which the
Hatree-Fock approximation is applied to the higher-order Green
function, the transport through the parallel-coupled DQD system
has been studied with an emphasis put on the effects of the
intra-dot Coulomb correlation on the Fano interference. We have
predicted that the Fano effect, as a consequence of quantum
interference, could survive in the presence of a finite Coulomb
repulsion. Without loss of generality, we select four states
corresponding to the bonding and antibonding DQD states and their
Coulomb blockade counterparts as the basis for investigation. The
coupling between each of these four states and the electrodes
broaden the level into a band with different width. The mechanism
of the Fano lineshape in conductance spectra is explored. It has
been also found that the direction of the asymmetric tail of Fano
lineshape can be flipped by tuning the dot-lead couplings. More
interestingly, by applying a magnetic flux, linear responses of
the four states could be interchanged, leading to a magnetic flux
tunable Fano effect. When a suitable bias voltage is applied, the
Fano peak doublet and a step-like lineshape may be observed as a
result of the splitting of equilibrium Fano resonance.

\begin{acknowledgments}
We would like to acknowledge Hui Zhai, Zuo-zi Chen, Chao-xing Liu,
and Zhen-Gang Zhu for helpful discussions. This work is supported
by  the Natural Science Foundation of China (Grant No. 10374056,
10574076), the MOE of China (Grant No. 2002003089), and the
Program of Basic Research Development of China (Grant No.
2001CB610508).

\end{acknowledgments}

\end{document}